\documentclass[11pt]{revtex4}
\usepackage{color}
\usepackage{amsmath,amssymb}
\usepackage[dvipdfm]{graphicx}
\usepackage{cancel}

\allowdisplaybreaks

\begin{document}
\title{A Higgs Quadruplet for Type III Seesaw
and \\Implications for $\mu \to e\gamma$ and $\mu - e$ Conversion}

\author{Bo Ren$^{1}$, Koji Tsumura$^2$, Xiao-Gang He$^{1,2}$, }

\affiliation{
$^1$INPAC, Department of Physics, Shanghai Jiao Tong University, Shanghai, China\\
$^2$Department of Physics and  Center for Theoretical Sciences, \\
National Taiwan University, Taipei, Taiwan
}

\date{\today $\vphantom{\bigg|_{\bigg|}^|}$}

\begin{abstract}
In Type III seesaw model the heavy neutrinos are contained in leptonic triplet representations. The Yukawa couplings of the triplet fermion and the left-handed neutrinos with the doublet Higgs field produce the Dirac mass terms. Together with the Majorana masses for the leptonic triplets, the light neutrinos obtain non-zero seesaw masses. We point out that it is also possible to have a quadruplet Higgs field to produce the Dirac mass terms to facilitate the seesaw mechanism. The vacuum expectation value of the quadruplet Higgs is constrained to be small by electroweak precision data. Therefore the Yukawa couplings of a quadruplet can be much larger than those for a doublet. We also find that unlike the usual Type III seesaw model where at least two copies of leptonic triplets are needed, with both doublet and quadruplet Higgs representations, just one leptonic triplet is possible to have a phenomenologically acceptable model because light neutrino masses can receive sizable contributions at both tree and one loop levels. Large Yukawa couplings of the quadruplet can induce observable effects for lepton flavor violating processes $\mu \to e \gamma$ and $\mu - e$ conversion. Implications of the recent $\mu \to e\gamma$ limit from MEG and also limit on $\mu - e $ conversion on Au are also given. Some interesting collider signatures for the doubly charged Higgs boson in the quadruplet are discussed.
\end{abstract}

\maketitle

\section{Introduction}

The type III seesaw contains leptonic triplets $\Sigma_R$ under the standard model (SM) gauge group $SU(3)_C\times SU(2)_L\times U(1)_Y$ as $(1,3,0)$,
$\Sigma_R = (\Sigma^+_R, \Sigma^0_R, \Sigma^-_R)$\cite{Ref:TypeIII}. In tensor notation, the triplet can be written as $\Sigma_R = (\Sigma_{ij})$ symmetric in $i$ and $j$,
where $i$ and $j$ take the values $1$ and $2$.
$\Sigma_{R11} = \Sigma_R^+$, $\Sigma_{R12} = i\Sigma_{R}^0/\sqrt{2}$ and $\Sigma_{R22} = \Sigma_{R}^-$. The Yukawa couplings related to neutrino and charged lepton masses come from the following terms
\begin{align}
L = -\bar L_L Y_e E_R \Phi -\bar L_LY_\nu \Sigma_{R} \tilde \Phi
- \frac12 \bar \Sigma^c_R M_R^\dagger \Sigma_R + H.c.
\end{align}
where the super-script ``c'' indicates the charge conjugation. The lepton doublet $L_L = (L_{Li}): (1,2,-1/2)$, $E_R = (E_{R_i}): (1,1,-1)$, and Higgs doublet $\Phi =( \phi_i): (1,2,1/2)$ ($\tilde \Phi = i\sigma_2 \Phi^*$) have the components given by $L_{L1} = \nu_L$, $L_{L2} = e_L$, and $\phi_1 = h^+$, $\phi_2 = (v+  h+i I_\phi)/\sqrt{2}$. With just one Higgs doublet, $I_\phi$ and $h^+$ are the would-be Nambu-Goldstone bosons $h_z$ and $h^+_w$ ``eaten'' by $Z$ and $W$ bosons, respectively.
We have
\begin{align}
&\bar L_L \Sigma_R \tilde \Phi
= \bar L_{Li} \Sigma_{Rij} \tilde \Phi_{j'}\epsilon^{jj'}
= - \Bigl(i {1\over 2 } \bar \nu_L\Sigma^0_R + {1\over \sqrt{2}}\bar e_L \Sigma_R^-\Bigr) ( v + h - i I_\phi)
- \Bigl(\bar \nu_L \Sigma^+_R + i {1\over \sqrt{2}} \bar e_L \Sigma^0_R\Bigr)
h^-\;,\nonumber\\
&\bar \Sigma^c_R \Sigma_R = \bar \Sigma^c_{Rij} \Sigma_{Ri'j'}\epsilon^{ii'}\epsilon^{jj'}
= \bar \Sigma^{-c}_R \Sigma_R^+ + \bar \Sigma^{0c}_R \Sigma^0_R + \bar \Sigma^{+c}_R \Sigma^-_R\;.
\end{align}
In the above, repeated indices are summed over from 1 to 2. $\epsilon_{12} = 1$, $\epsilon_{21} = -1$ and $\epsilon_{11}=\epsilon_{22} = 0$.
The neutrino and charged lepton mass matrices $M_\nu$ and $M_E$, in the basis $(\nu_L^c, \Sigma^{0}_R)^T$ and $(e_R, \Sigma^-_R)^T$, are given by
\begin{align}
M_\nu = \left ( \begin{array}{cc}
0& M_{\nu\Sigma}\\
M_{\nu\Sigma}^T &M_R^\dagger
\end{array}\right )\;,\;\;\;\;
M_E = \left ( \begin{array}{cc}
M_e& M_{e\Sigma}\\
0 & M_R^\dagger
\end{array}\right )\;,
\end{align}
where Dirac mass term $M_{\nu \Sigma} = - i Y_\nu v/2 $, $M_{e\Sigma} = -  Y_\nu v/\sqrt{2}$ and $M_e = Y_e v/\sqrt{2}$ where $v$ is the vacuum expectation value (VEV) of the Higgs doublet.

Note that given $L_L$ and $\Sigma_R$ representations, it is also possible to have the necessary Dirac mass term $M_{\nu \Sigma}$ from the Yukawa couplings of a quadruplet Higgs representation $\chi$: $(1,4,-1/2)$ of the following form,
\begin{align}
L = - \bar L_L Y_\chi \Sigma_R \chi + H.c.
\end{align}
The field $\chi$ has component fields: $\chi = (\chi^+, \chi^0, \chi^-, \chi^{--})$. In tensor notation $\chi$ is a total symmetric tensor with 3 indices $\chi_{ijk}$
with $i$, $j$ and $k$ taking values $1$ and $2$ with
\begin{align}
\chi_{111} = \chi^+\;,\;\;
\chi_{112} = {1\over \sqrt{3}}\chi^0\;,\;\;
\chi_{122} = {1\over \sqrt{3}}\chi^-\;,\;\;
\chi_{222} =\chi^{--}\;.
\end{align}
We have
\begin{align}
\bar L_L \Sigma_R \chi
 &= \bar L_{Li} \Sigma_{Rjk} \chi_{ij'k'}\epsilon^{jj'}\epsilon^{kk'}\nonumber\\
 &= \bar \nu_L \Bigl({1\over \sqrt{3}} \Sigma^+_R \chi^-
- i \sqrt{{2\over 3}}\Sigma^0_R \chi^0 + \Sigma^-_R\chi^+ \Bigr)
+  \bar e_L \Bigl(\Sigma^+_R \chi^{--} - i\sqrt{{2\over 3}}\Sigma^0_R \chi^-
+ {1\over \sqrt{3}} \Sigma^-_R\chi^0\Bigr)\;.   \label{mixing}
\end{align}
The neutral component $\chi^0$ can have VEV $v_\chi$ with $\chi^0 = (v_\chi + \chi_R + i \chi_I)/\sqrt{2}$.
A non-zero $v_\chi$ will modify the neutrino and charged lepton mass matrices $M_{\nu \Sigma}$ and $M_{e\Sigma}$ with
\begin{align}
M_{\nu \Sigma} = - i{1\over 2}Y_\nu v -i {1\over \sqrt{3}} Y_\chi v_\chi\;,\;\;\;
\;M_{e\Sigma} = -{1\over \sqrt{2}}Y_e v + {1\over \sqrt{6}} Y_\chi v_\chi\;.
\end{align}
To the leading tree level light neutrino mass matrix $m_\nu$, defined by $L_m = - \frac12\bar \nu^c_L m_\nu\nu_L$ +H.c., is given by
\begin{align}
 m_\nu = - M^*_{\nu \Sigma} M_R^{-1}M^\dagger_{\nu \Sigma}
= \Bigl({1\over 2} Y_\nu^* v + {1\over \sqrt{3}} Y^*_\chi v_\chi \Bigr)
M^{-1}_R
\Bigl({1\over 2} Y_\nu^\dagger v + {1\over \sqrt{3}} Y^\dagger_\chi v_\chi \Bigr)\;.
\label{tree}
\end{align}
A model with a different Higgs quadruplet $(1,4,3/2)$ has also been studied where
neutrino masses only arises from a dimension-7 operator \cite{Ref:dim7}. This model is very different from the model we are discussing here.

In the basis where the charged lepton mass matrix is already diagonalized, the PMNS mixing matrix $V$\cite{Ref:Pontecorvo,Ref:MNS} in the charged current interaction
is given by
\begin{align}
\hat m_\nu = V^T m_\nu V\;,
\end{align}
where $\hat m_\nu = diag(m_1, m_2, m_3)$ is the diagonalized light neutrino mass matrix.

The introduction of quadruplet $\chi$ in the model can have interesting consequences for  neutrino masses, mixing and also for lepton flavor violating (LFV) processes,
$\mu \to e \gamma$ and $\mu - e$ conversion because the VEV of $\chi$ is constrained to be small which then can lead to a large Yukawa coupling $Y_\chi$. We also found some interesting collider signatures of the doubly charged Higgs boson in the quadruplet. In the following we will study the quadruplet model in more details.

\section{The electroweak constraint}

We have seen that in Type III seesaw, it is possible to introduce a quadruplet Higgs which give additional seesaw  contributions
to neutrino masses at the tree level. It is, however, well known that electroweak precision data constrain the VEV of a
Higgs representation because  a non-zero VEV of some Higgs may break the $SU(2)$ custodial symmetry in the SM
leading to a large deviation of the $\rho$ parameter from unity. With
the constraints satisfied, the Higgs doublet and quadruplet may contribute to the neutrino mass matrix differently.

The non-zero VEV of the Higgs representation with isospin $I$ and hypercharge $Y$ will modify the $\rho$ parameter at tree level with\cite{Ref:HHG},
\begin{align}
\rho = {\sum_\alpha (I_\alpha (I_\alpha + 1) - Y_\alpha^2)v^2_\alpha
\over \sum_\alpha 2 Y^2_\alpha v^2_\alpha}\;.
\end{align}
The SM doublet Higgs alone does not lead to a deviation  of $\rho$ from unity, but the addition of a quadruplet does.
For our case of one doublet and one quadruplet, we have
\begin{align}
\rho = {v^2 + 7 v^2_\chi\over v^2+ v^2_\chi}
= 1 +{6v^2_\chi\over v^2+v^2_\chi}\;.
\end{align}
We therefore have, $\Delta \rho = 6 v^2_\chi/(v^2+v^2_\chi) =6\sqrt2 G_F v_\chi^2$.
Using experimental data
$\Delta \rho  = 0.0004^{+0.0029}_{-0.0011}$(95\% c.l.)\cite{Ref:PDG}, we see that $v_\chi$ is constrained to be less than 5.8 GeV which is about 40 times smaller than that of the doublet Higgs VEV. This vast difference in Higgs VEV's indeed indicate that the Higgs doublet and quadruplet contribute to the neutrino mass matrix differently in the sense that if the Yukawa couplings $Y_\nu$ and $Y_\chi$ are the same order of magnitude, they contribute to the neutrino masses can be different by orders of magnitude. Turning this around, if both Higgs contribute to the neutrino masses with the same orders of magnitude, then the Yukawa coupling for quadruplet $Y_\chi$ can be several orders of magnitude larger than
that for the doublet $Y_\nu$.

If the seesaw mass is only from the coupling to $\Phi$, just like Type III seesaw with one doublet, the canonical Yukawa coupling is of order $\sqrt{M_R m_\nu/v^2}$. With a $M_R$ of order $1$ TeV, the Yukawa couplings would be less than
$10^{-5}$ with $m_\nu$ around $0.1$ eV. This makes it clear that even the heavy degrees of freedom are kinematically accessible at the LHC\cite{Ref:type3lhc}, the small Yukawa couplings is hard to study their properties and their effects on LFV processes\cite{Ref:type3lfv1,Ref:type3lfv2,Ref:type3lfv4}. Although it has been shown that there are solutions with large Yukawa coupling in Type III seesaw with just one Higgs doublet\cite{Ref:type3lfv2,Ref:type3lfv3}, it is interesting to see if large Yukawa couplings can more naturally manifest itself.
The quadruplet with a small VEV provides such a
possibility. The natural size of the Yukawa coupling $Y_\chi$ is of order $\sqrt{M_R m_\nu/v_\chi^2}$. With $v_\chi$ of order 1 GeV,
$Y_\chi$ would be enhanced by about 250 times compared with $Y_\nu$. With a smaller $v_\chi$, $Y_\chi$ can be even larger since
$Y_\chi \sim 10^{-3}(1\mbox{GeV}/v_\chi) \sqrt{(M_R/\mbox{TeV}) (m_\nu/0.1\mbox{eV})}$. The large Yukawa coupling $Y_\chi$ can
lead to interesting phenomenology, such as the possibility of having
large effects in lepton flavor violating (LFV) processes $\mu \to e\gamma$ and $\mu - e$ conversion.

\section{Loop induced neutrino mass with just one triplet lepton}
In the Type III seesaw with just doublet Higgs, if there is just one leptonic triplet $\Sigma_R^{}$, the resulting neutrino mass
matrix $m_\nu$ for the three light neutrinos is only a rank one matrix. This implies that only one light neutrino mass is non-zero.
Neutrino oscillation data show the existence of two distinct mass
squared splittings, so a model with just one generation of triplet $\Sigma_R^{}$ is in conflict with data.
More than one generation of $\Sigma_R^{}$ is required to have a higher ranked mass matrix to fit data.
We point out that with the introduction of quadruplet $\chi$, it is possible to raise the rank of neutrino mass matrix by including one loop
contributions to the mass matrix. The tree and loop generated mass matrices together can be consistent with present data on neutrino mass and mixing.
With both Higgs doublet and quadruplet, the tree level light neutrino mass matrix $m_\nu$ given in eq.\eqref{tree} is still rank one if there is only one generation of $\Sigma_R^{}$. In the following we show that the inclusion of one loop contribution can raise the rank of the mass matrix to two.

The one loop contributions involve exchange of internal quadruplet Higgs bosons and heavy leptons.
In order to show this mechanism explicitly, we first identify physical Higgs states and mixing necessary for one loop generation of neutrino mass from the Higgs potential.
The most general renormalizable Higgs potential is given by
\begin{align}
{\mathcal V}
&= -\mu^2 \left( \Phi^\dag\Phi \right)
+\lambda \left( \Phi^\dag\Phi \right)^2
+M^2\chi^\dag\chi
+\lambda_\chi^\alpha (\chi^\dag \chi \chi^\dag \chi)_\alpha
+\lambda_{\Phi\chi}^\alpha (\Phi^\dag \Phi \chi^\dag \chi)_\alpha \nonumber \\
&\qquad
+\left [ \frac{\lambda_5}{2}(\Phi\chi)^2 +  \lambda_{3\Phi} \Phi^\dagger \Phi \Phi \chi + \text{H.c}\right],
\end{align}
where $\alpha$ denotes an index for $SU(2)$ contractions.
The contraction of $SU(2)$ indices for each of the terms are given by
\begin{align}
\chi^\dag \chi &= \chi_{ijk}^* \chi_{ijk}, \nonumber\\
(\chi^\dag \chi \chi^\dag \chi)_1
&= \chi_{ijk}^* \chi_{ijk} \chi_{lmn}^* \chi_{lmn},\nonumber \\
(\chi^\dag \chi \chi^\dag \chi)_2
&= \chi_{ijk}^* \chi_{ijn} \chi_{lmn}^* \chi_{lmk}, \nonumber\\
(\chi^\dag \chi \chi^\dag \chi)_3
&= \chi_{ijk}^* \chi_{rjk} \chi_{lmn}^* \chi_{smn} \epsilon_{il} \epsilon_{rs}, \nonumber\\
(\chi^\dag \chi \chi^\dag \chi)_4
&= \chi_{ijk}^* \chi_{rsk} \chi_{lmn}^* \chi_{tun}
\epsilon_{il} \epsilon_{jm} \epsilon_{rt} \epsilon_{su},\\
(\Phi^\dag \Phi \chi^\dag \chi)_1 &= \Phi_i^* \Phi_i \chi_{jkl}^* \chi_{jkl}, \nonumber\\
(\Phi^\dag \Phi \chi^\dag \chi)_2 &= \Phi_i^* \Phi_j \chi_{jkl}^* \chi_{ikl}, \nonumber \\
(\Phi^\dag \Phi \chi^\dag \chi)_3
&= \Phi_i^* \Phi_j \chi_{klm}^* \chi_{nlm} \epsilon_{ik} \epsilon_{jn},\nonumber \\
(\Phi \chi)^2
&= \Phi_i \Phi_j \chi_{i'kl} \chi_{j'mn} \epsilon_{ii'}\epsilon_{jj'}\epsilon_{km} \epsilon_{ln},\nonumber \\
\Phi^\dagger \Phi \Phi \chi &= \Phi^*_i\Phi_j\Phi_k\chi_{ij'k'}\epsilon_{jj'}\epsilon_{kk'}\;.\nonumber
\end{align}
In the above only two terms are independent for $(\chi^\dagger\chi\chi^\dagger \chi)_\alpha$.
Also only two terms are independent for  $(\Phi^\dagger \Phi\chi^\dagger \chi)_\alpha$. One can just take $\alpha$ to be equal to 1 and 2 as the independent terms for these two types of terms. In the following, we set $\lambda_\chi^3=\lambda_\chi^4=\lambda_{\Phi\chi}^3=0$ without loss of generality.

The two terms $(\Phi\chi)^2$ and $\Phi^\dagger \Phi\Phi \chi$, break the global lepton number
symmetry after the doublet and quadruplet develop non-zero VEV's.  $\Phi^\dagger \Phi\Phi \chi$ then mixes $\Phi$ and $\chi$ fields.
At one loop level Majorana masses will be generated for light neutrinos. There are three types of mixing
terms which can be characterized to be proportional to $v^2$, $vv_\chi$ or $v_\chi^2$. We have seen earlier that $v$ is much
larger than $v_\chi$ from electroweak precision data, therefore
one can just keep terms proportional to $v^2$ for the loop generation of neutrino masses. These terms are
\begin{align}
L &= - {1\over 2} \lambda_5 v^2 \Bigl( {1\over \sqrt{3}}\chi^+\chi^- - {1\over 6} (\chi_R + i \chi_I)^2 \Bigr) \nonumber \\
&\qquad -v^2\lambda_{3\Phi} \Bigl[ \Bigl({1\over 2} h^- \chi^+
- {1\over \sqrt{3}} h^+ \chi^- \Bigr)
+{1\over 4\sqrt{3}}(3 h+i I_\phi)(\chi_R +i \chi_I) \Bigr] + H.c.
\end{align}
The above terms will generate a neutrino mass matrix proportional to $Y^*_\chi Y_\chi^\dagger$ for the
first term and, $Y^*_\nu Y^\dagger_\chi$ for the second term.
To have a consistent model, the elements in $Y_\nu$ are required to be much smaller than those in $Y_\chi$.
We can neglect the contribution from terms proportional to $\lambda_{3\Phi}$ in the above.
Without terms proportional to $\lambda_{3\Phi}$ and $v_\chi$, masses of component fields in $\chi$ are given by
\begin{align}
m_{\chi_R}^2 &\simeq  M^2 +\Bigl( \frac12 \lambda_{\Phi\chi}^1
+\frac16\lambda_{\Phi\chi}^2 -\frac13\lambda_5 \Bigr)v^2, \nonumber\\
m_{\chi_I}^2 &\simeq  M^2 +\Bigl( \frac12 \lambda_{\Phi\chi}^1
+\frac16\lambda_{\Phi\chi}^2 +\frac13\lambda_5 \Bigr) v^2, \\
m_{\chi^{\pm\pm}}^2 &\simeq  M^2+\frac12 \left( {\lambda_{\Phi\chi}^1+\lambda_{\Phi\chi}^2} \right) v^2.\nonumber
\end{align}
We note that a parameter $\lambda_5$ characterizes
a mass squared splitting between $\chi_R$ and $\chi_I$, i.e., $(m_{\chi_R}^2-m_{\chi_I}^2)\simeq - (2/3) \lambda_5 v^2$.
The mass matrix for singly charged scalars is given by
\begin{align}
&\begin{pmatrix} \chi^{+*} & \chi^- \end{pmatrix}
\begin{pmatrix}
M^2 +\tfrac12 \lambda_{\Phi\chi}^1v^2
& \tfrac1{2\sqrt3} \lambda_5 v^2 \\ \tfrac1{2\sqrt3} {\lambda_5} v^2
& M^2 +\tfrac12 [\lambda_{\Phi\chi}^1 +\tfrac23\lambda_{\Phi\chi}^2]v^2
\end{pmatrix}
\begin{pmatrix} \chi^+ \\ \chi^{-*} \end{pmatrix} =
\begin{pmatrix} \chi_1^- & \chi_2^- \end{pmatrix}
\begin{pmatrix} m_{\chi^\pm_1}^2 & \\ & m_{\chi^\pm_2}^2 \end{pmatrix}
\begin{pmatrix} \chi_1^+ \\ \chi_2^+ \end{pmatrix}\;,
\end{align}
where
\begin{align}
\begin{pmatrix} \chi_1^+ \\ \chi_2^+ \end{pmatrix} =
\begin{pmatrix} \cos\theta & -\sin\theta \\ \sin\theta & \cos\theta \end{pmatrix}
\begin{pmatrix} \chi^+ \\ \chi^{-*} \end{pmatrix}
\quad \text{ with } \quad
\tan2\theta =  -\frac{\sqrt3\, \lambda_5}{\lambda_{\Phi\chi}^2}.
\end{align}

The one-loop contributions to the neutrino mass matrix are calculated as\cite{Ref:MaModel,Ref:MaSuematsuModel}
\begin{align}
M_\nu^\text{loop}
&\simeq Y_\chi^*Y_\chi^\dagger {1\over 8\pi^2} \Biggl\{ {1\over 3}
m_N \Bigl[ I \Bigl(\frac{m^2_{\chi_R}}{m^2_N}\Bigr)
- I\Bigl(\frac{m^2_{\chi_I}}{m^2_N}\Bigr)\Bigr]
- {\sin(2\theta)\over  \sqrt{3}}
m_E \Bigl[ I \Bigl(\frac{m^2_{\chi^+_1}}{m^2_E} \Bigr)
- I \Bigl( \frac{m^2_{\chi^+_2}}{m^2_E}\Bigr) \Bigr] \biggr\}\;,
\end{align}
where $m_E$ and $m_N$ are masses of neutral and charged heavy leptons, and
$I(x) = x\ln x/ (1-x)$.
The explicit dependence on $\lambda_5$ is given
\begin{align}
&M_\nu^\text{loop}
\simeq {\kappa\over m_N}  Y_\chi^*Y_\chi^\dagger v^2\;, \nonumber\\
&\kappa = {\lambda_5 \over 12\pi^2} \Bigl\{
-\frac13 {m^2_N\over m^2_{\chi_R} - m^2_{\chi_I}}
\Bigl[ I\Bigl(\frac{m^2_{\chi_R}}{m^2_N}\Bigr)
-I\Bigl(\frac{m^2_{\chi_I}}{m^2_N}\Bigr)\Bigr]
+ \frac12{m_N m_E\over m^2_{\chi^+_1} - m^2_{\chi^+_2}}
\Bigl[ I\Bigl(\frac{m^2_{\chi^+_1}}{m^2_E}\Bigr)
- I\Bigl(\frac{m^2_{\chi^+_2}}{m^2_E}\Bigr)\Bigr] \Bigr\}\;.
\end{align}
Neglecting mass splitting in a multiplet, i.e., $m_{\chi_i} = m_\chi$, $m_N^{} = m_E^{}$, $\kappa$ is given by
\begin{align}
\kappa = {\lambda_5 \over 72 \pi^2} J(m^2_\chi/ m^2_N)\;,\;\; J(x) = {1\over 1-x} + {\ln x\over (1-x)^2}\;,
\end{align}
where $J(1)=-1/2$.

Collecting contributions from the tree and loop contribution, one can write the neutrino mass matrix as
\begin{align}
M_\nu^{ij}
= \left( M_\nu^\text{tree} +M_\nu^\text{loop} \right)^{ij}
= {1\over m_N}
\Bigl({1\over 2} Y_\nu^i v + {1\over \sqrt{3}} Y_\chi^i v_\chi\Bigr)^*
\Bigl({1\over 2} Y^j_\nu v + {1\over \sqrt{3}} Y_\chi^j v_\chi \Bigr)^*
+{\kappa \over m_N}  Y_\chi^{i*} Y_\chi^{j*}v^2.
\label{Eq:rank2}
\end{align}
The mass matrix is now rank 2 in general. This mechanism can also work even when we introduce an additional scalar doublet\cite{Ref:2hdmTreeLoop}.
However such a scalar is indistinguishable from a SM Higgs doublet without additional quantum charges. The
extra doublet fields can interact with other SM fermions and will induce large tree level flavor changing neutral current (FCNC) for the charged leptons. In this model, the tree level FCNC are much suppressed
for charged leptons.

\section{Some Phenomenological implications}

\subsection{Neutrino masses and mixing}
The mass matrix obtained in the previous section, being rank two, has two non-zero eigenvalues. One of the neutrino masses is predicted to be zero. The zero mass neutrino can be $m_{\nu_1}$ or $m_{\nu_3}$ depending on whether the neutrino masses have normal or inverted hierarchy. In this section, we show that the mass matrix obtained can be made consistent with experimental data on mixing parameters.

Mass squared differences of neutrino masses and neutrino mixing have been measured to good precision\cite{Ref:dm12,Ref:dm31,Ref:th23,Ref:th13,Ref:th12,Ref:t2k}.
The mass parameters are determined by global fit as\cite{Ref:nuparam}
$\Delta m_{21}^2 = (7.58^{+0.22}_{-0.26}) \times 10^{-5}$ eV${}^2$, $|\Delta m_{31}^2| = (2.39^{+0.12}_{-0.09}) \times 10^{-3}$ eV${}^2; (2.31^{+0.12}_{-0.09}) \times 10^{-3}$ eV${}^2)$ for normal (inverted) mass hiearchy. Here $\Delta m^2_{ij} = m_i^2-m_j^2$, and $m_i(i=1$--$3)$. For our case, with normal hierarchy,  $m^2_1 = 0$, $m_2^2 = \Delta m_{12}^2$ and $m^2_3 = \Delta m_{31}^2$. For inverted hierarchy, we then have $m_3^2 = 0$, $m_1^2 = -\Delta m^2_{31}$, and $m^2_2 = \Delta m^2_{21} - \Delta m^2_{31}$.
The neutrino mixing are given by\cite{Ref:nuparam} $\sin^2(\theta_{23})=0.42_{-0.03}^{+0.08}$, $\sin^2(\theta_{12})=0.306_{-0.005}^{+0.018}$, and $\sin^2(\theta_{13})<0.028$.
To the leading order, the mixing pattern can be approximated by the tribimaximal mixing matrix\cite{Ref:TB},
\begin{align}
U_\text{TB} =
\begin{pmatrix}
\sqrt{\frac23} & \frac1{\sqrt{3}} & 0 \\
-\frac1{\sqrt{6}} & \frac1{\sqrt{3}} & \frac1{\sqrt2} \\
-\frac1{\sqrt{6}} & \frac1{\sqrt{3}} & -\frac1{\sqrt2}
\end{pmatrix}.
\end{align}
The light neutrino mass matrix obtained in eq.\eqref{Eq:rank2} can be easily made to fit data. We consider
the case for $v \gg v_\chi$, such that terms proportional to $v_\chi$ can all be neglected for illustration.
With this approximation, the cross term proportional to $Y^*_\nu Y_\chi^\dagger  + Y_\chi^* Y_\nu^\dagger$ can be neglected.

For normal hierarchy case, by imposing the condition of the tribimaximal mixing, the Yukawa couplings can be taken to be the forms $Y_\nu \sim y_\nu (0,\;1/\sqrt{2},\;-1/\sqrt{2})^T$,
 and $Y_\chi \sim y_\chi(1/\sqrt{3},\;1/\sqrt{3},\;1/\sqrt{3})^T$. In this case
$m_3 = y^2_\nu v^2/4 m_N $ and $m_2=\kappa y_\chi^2 v^2/m_N$. If the heavy neutrino mass is of order 1 TeV, $y_\nu \sim 1.80\times 10^{-6} (m_N/1 \text{TeV})^{1/2}$ and $\sqrt{\kappa} y_\chi = 0.38\times 10^{-6} (m_N/1 \text{TeV})^{1/2}$. We note that relative size of tree level and loop level contributions can be tuned by the parameter $\kappa$, which is proportional to the Higgs potential parameter $\lambda_5$. If $\lambda_5$ is small, quaruplet Yukawa coupling $y_\chi$ can be order of one.
This kind of possibility is also studied in the neutrinophilic two Higgs doublet model\cite{Ref:HT}. The role of the $Y_\nu$ and $Y_\chi$ can be switched.

Similarly the model can be made consistent with inverted hierarchy. For example with $Y_\nu \sim y_\nu (\sqrt{2/3},\;-1/\sqrt{6},\;-1/\sqrt{6})^T$ and $Y_\chi \sim y_\chi(1/\sqrt{3},\;1/\sqrt{3},\;1/\sqrt{3})^T$, the tribimaximal mixing pattern can be realized.
In this case $m_1 = y^2_\nu v^2/4 m_N $ and $m_2=\kappa y_\chi^2 v^2/m_N$. If the heavy neutrino mass is of order 1 TeV, $y_\nu \sim 1.78\times 10^{-6} (m_N/1 \text{TeV})^{1/2}$ and $\sqrt{\kappa} y_\chi = 0.90\times 10^{-6} (m_N/1 \text{TeV})^{1/2}$. Again the roles of $Y_\nu$ and $Y_\chi$ can be switched.

Making perturbation to the above forms, one can get non-zero $\theta_{13}$ solutions, which is indicated by recent results at T2K\cite{Ref:t2k}. For instance, for normal mass hierarchy case modifying $Y_\nu$ to be $Y_\nu^{'}=Y_\nu+\Delta Y_\nu = Y_\nu+ y_\nu(a, b, c)^T$ and keep the same $Y_\chi$, we can produce non-zero $\theta_{13}$ solutions. Using the $\Delta Y_\nu = y_\nu(-0.14, 0, 0)^T, y^2_\nu v^2/4 m_N =5.23\times 10^{-2}$ eV, $\kappa y_\chi^2 v^2/m_N=9.14\times 10^{-3}$eV, we obtain $m_2= 8.78\times10^{-3}$, $m_3 = 4.82\times10^{-2}$, $\sin^2\theta_{12} = 0.323$, $\sin^2\theta_{23} = 0.44$ and $\sin^2\theta_{13} = 0.025$ which are within one $\sigma$ error of the data.

For inverted mass hierarchy case, with $\Delta Y_\nu = y_\nu(-0.0095, -0.1, 0.1085)^T, y^2_\nu v^2/4 m_N =4.81\times 10^{-2}$eV, $\kappa y_\chi^2 v^2/m_N=4.88 \times 10^{-2}$eV, we obtain $m_1= 4.80\times10^{-2}$, $m_2 = 4.88\times10^{-2}$, $\sin^2\theta_{12} = 0.306$, $\sin^2\theta_{23} = 0.41$ and $\sin^2\theta_{13} = 0.014$ which are, again,  within one $\sigma$ error of the data.

Higher order loop corrections can further raise the rank of neutrino mass matrix in general. Therefore, all three light neutrinos can have non-zero masses in this model.
It has been shown in Ref.\cite{Ref:2loop} that the rank of the neutrino mass matrix can be rank two at two loop level even with just one triplet lepton and one Higgs doublet. However, in this case the heavy triplet lepton mass needs to be $10^{16}$ GeV, and hence its phenomenological
consequence for collider physics is out of the scope at the LHC.
Introduction of more leptonic triplet generations can also increase the rank of mass matrix too.

\subsection{$\mu \to e\gamma$ and $\mu - e$ conversion}
We now study possible effects on LFV processes $\mu \to e \gamma$ and $\mu - e$ conversion. $\mu \to e \gamma$ is induced at one loop level.
There is a small contribution to $\mu - e $
conversion at the tree level due to mixing of charged light and heavy leptons.
The dominant contribution come at the one loop level due to possible
large Yukawa coupling $Y_\chi$, because the size of $Y_\nu$ is constrained to be small by the absolute size of neutrino masses and the doublet Higgs VEV.
The one loop induced effective Lagrangian responsible to $\mu \to e \gamma$ and $\mu - e$ conversions is given by
\begin{align}
{\mathcal L} &= - \overline{\psi_\mu} \sigma^{\mu\nu}
(A_L P_L +A_R P_R)\psi_e F_{\mu\nu}
+ \sum_q e Q_q \bar q \gamma^\mu q \overline{\psi_\mu} \gamma_\mu P_L \psi_e B_L + \text{H.c}\;,
\end{align}
with $Q_q$ being the electric charge of the $q$-quark, and
\begin{align}
A_L &= \frac{e}{32\pi^2} Y_\chi \Bigl\{
-\tfrac16 \bigl[ \tfrac1{m_{\chi_R}^2}F_\Sigma(\tfrac{m_E^2}{m_{\chi_R}^2})
+\tfrac1{m_{\chi_I}^2}F_\Sigma(\tfrac{m_E^2}{m_{\chi_I}^2}) \bigr]
+\tfrac23 \bigl[ \tfrac{s_\theta^2}{m_{\chi^+_1}^2}F_\chi(\tfrac{m_N^2}{m_{\chi^+_1}^2})
+\tfrac{c_\theta^2}{m_{\chi^+_2}^2}F_\chi(\tfrac{m_N^2}{m_{\chi^+_2}^2}) \bigr]
\nonumber \\
&\qquad +\tfrac1{m_{\chi^{++}}^2} \bigl[ F_\Sigma(\tfrac{m_E^2}{m_{\chi^{++}}^2})
+2F_\chi(\tfrac{m_E^2}{m_{\chi^{++}}^2}) \bigr]
\Bigr\} Y_\chi^\dag m_\mu,\nonumber\\
A_R &={m_e\over m_\mu} A_L,\\
B_L &= \frac{e}{16\pi^2} Y_\chi \Bigl\{
-\tfrac16 \bigl[ \tfrac1{m_{\chi_R}^2}G_\Sigma(\tfrac{m_E^2}{m_{\chi_R}^2})
+\tfrac1{m_{\chi_I}^2}G_\Sigma(\tfrac{m_E^2}{m_{\chi_I}^2}) \bigr]
+\tfrac23 \bigl[ \tfrac{s_\theta^2}{m_{\chi^+_1}^2}G_\chi(\tfrac{m_N^2}{m_{\chi^+_1}^2})
+\tfrac{c_\theta^2}{m_{\chi^+_2}^2}G_\chi(\tfrac{m_N^2}{m_{\chi^+_2}^2}) \bigr]
\nonumber \\
&\qquad +\tfrac1{m_{\chi^{++}}^2} \bigl[ G_\Sigma(\tfrac{m_E^2}{m_{\chi^{++}}^2})
+2G_\chi(\tfrac{m_E^2}{m_{\chi^{++}}^2}) \bigr]\Bigr\} Y_\chi^\dag\;,\nonumber
\end{align}
where
\begin{align}
&F_\Sigma(z) =
\frac{z^2-5z-2}{12(z-1)^3} +\frac{z\ln z}{2(z-1)^4}\;,\;\;\;F_\chi(z) =
\frac{2z^2+5z-1}{12(z-1)^3} -\frac{z^2\ln z}{2(z-1)^4}\;,\nonumber\\
&G_\Sigma(z) = {7z^3-36z^2 +45 z -16 +6(3z-2)\ln z\over 36(1-z)^4}\;,\;\;\;\;G_\chi(z) = {11z^3-18z^2+9z-2 - 6 z^3\ln z\over 36(1-z)^4}\;.
\end{align}
The LFV $\mu \to e \gamma$ decay branching ratio is easily evaluated by
\begin{align}
B(\mu \to e \gamma) = \frac{48\pi^2}{G_F^2m_\mu^2} (|A_L|^2 + |A_R|^2).
\end{align}
The strength of $\mu - e$ conversion is measured by the quantity, $B^A_{\mu \to e}= \Gamma^A_{conv}/\Gamma^A_{capt} = \Gamma(\mu^- + A(N,Z) \to e^- +A(N,Z))/\Gamma(\mu^- + A(N,Z) \to \nu_\mu + A(N+1, Z-1)$. Following Ref.\cite{Ref:Mu2Ecv1}, we have
\begin{align}
{B^A_{\mu\to e}\over B(\mu \to e\gamma)} = R^0_{\mu\to e}(A) \left | 1 + {\tilde g^{(p)}_{LV} V^{(p)}(A)\over A_R D(A)} + {\tilde g^{(n)}_{LV} V^{(n)}(A)\over A_R D(A)}\right |^2\;,
\end{align}
where
\begin{align}
R^0_{\mu\to e}(A) = {G^2_F m^5_\mu \over 192 \pi^2 \Gamma^A_{capt}}|D(A)|^2\;.
\end{align}
and $\tilde g^{(p)}_{LV} = 2 g_{LV(u)}+g_{LV(d)}, \tilde g^{(n)}_{LV} = g_{LV(u)}+ 2 g_{LV(d)}$ with $g_{LV(q)} =  -e Q_q B_L /(\sqrt2 G_F).$

\begin{figure}
\centering
\includegraphics[width=6.8cm]{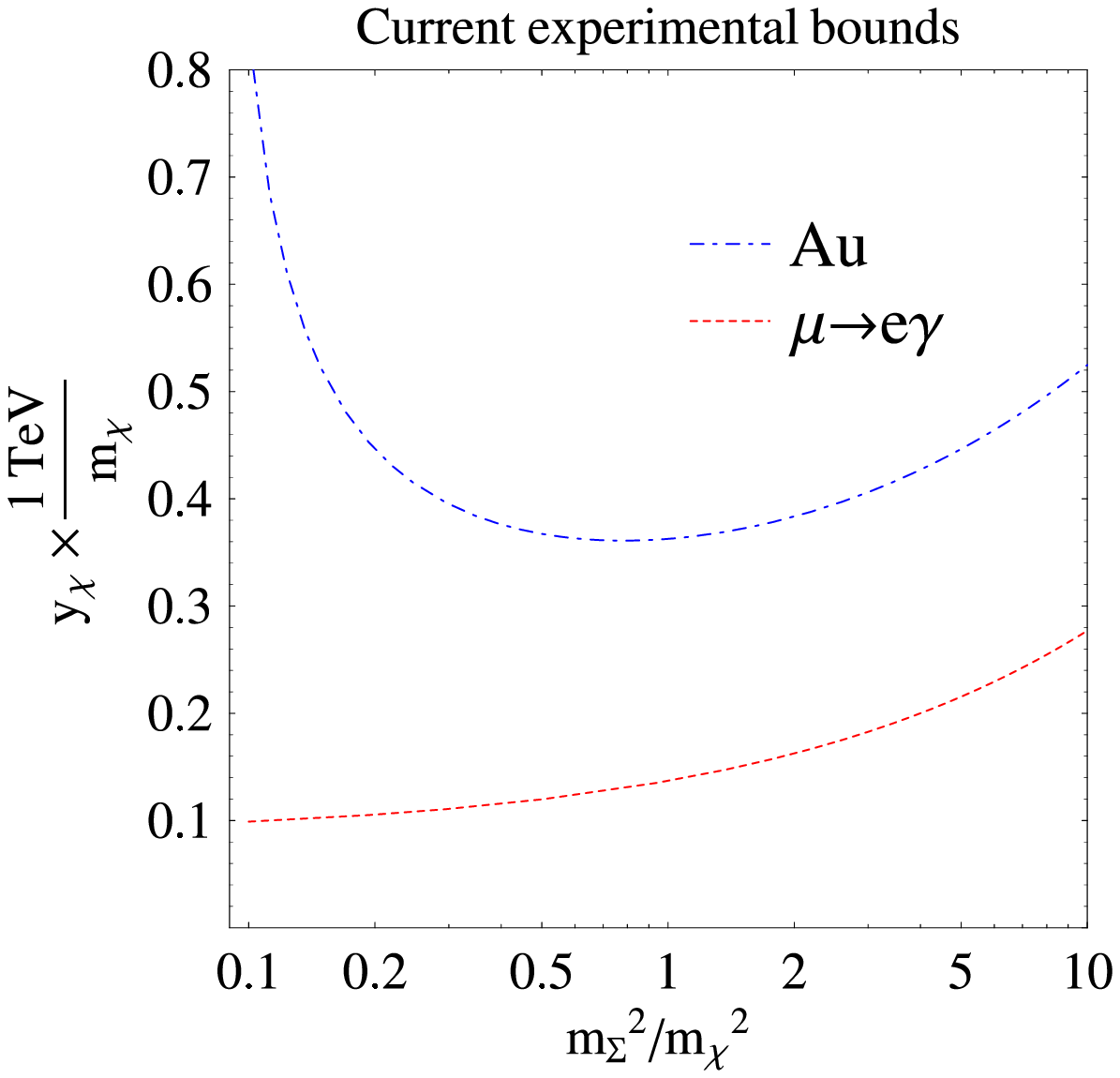}
\includegraphics[width=7cm]{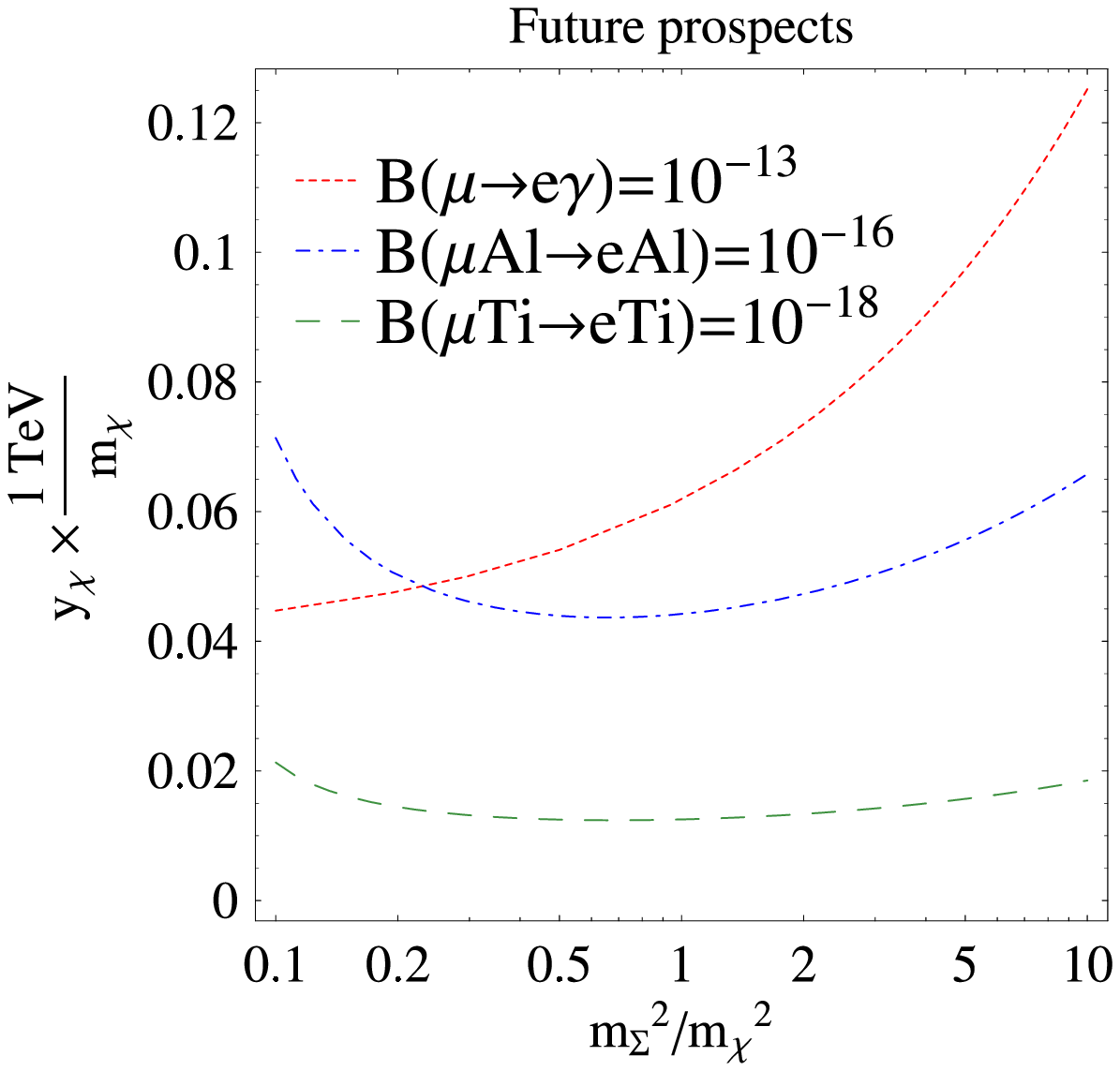}
\caption{The current and future experimental constraints on the quadruplet Yukawa couplings from $\mu \to e\gamma$ and $\mu-e$ conversion. The mass of quadruplet scalar is taken as $m_\chi = 1$ TeV. }
\label{FIG:LFV}
\end{figure}

For many years, the best 90\% c.l. experimental upper limit for  $B(\mu \to e\gamma)$ was $1.2\times 10^{-11}$\cite{Ref:MEGA}. Recently, MEG collaboration has obtained better result with the 90\% c.l. upper limit\cite{meg-new} $2.4\times 10^{-12}$. This new bound, as will be seen, provides important constraint for the quadruplet model discussed here.
There are several measurements of $\mu - e$ conversion on various nuclei. The best bound is for Au nuclei with the 90\% c.l. experimental bound given by $B^{\text{Au}}_{\mu \to e} < 7\times 10^{-13}$\cite{Ref:Mu2E-Au}.  For Au, the relevant parameters determined by method I in Ref.\cite{Ref:Mu2Ecv1} are given by: $D(\text{Au})=0.189$, $V^{(p)}(\text{Au})=0.0974, V^{(n)}(\text{Au})=0.146$ and $R^0_{\mu\to e}(\text{Au})=0.0036$\cite{Ref:Mu2Ecv1}. We will use these values to study implication for our quadruplet model.

The numerical results are shown in FIG.\ref{FIG:LFV}. In obtaining results in FIG.\ref{FIG:LFV}, we have choosen the mass of quadruplet component field $\chi_i$ to be degenerate with a common mass of 1 TeV, and the quadruplet Yukawa coupling constant is taken as $Y_\chi \sim y_\chi(1/\sqrt{3},\;1/\sqrt{3},\;1/\sqrt{3})^T$ which satisfying neutrino mixing data from our previous studies for illustration. In the left panel of FIG.\ref{FIG:LFV}, we show current experimental bounds on the quadruplet Yukawa coupling from non-observation of $\mu\to e\gamma$ and $\mu -e$ conversion as a function of ratio of triplet fermion and quadruplet scalar squared masses.
We found that current constraints on quadruplet Yukawa coupling constant from $\mu-e$ conversions are weaker than that from $\mu\to e\gamma$. This is very different than the situation in a model with fourth generation where non-zero $Z$-penguin contribution dominates and $\mu - e $ conversion gives stronger constraints\cite{Ref:meco4g}. In the quadruplet model discussed here because the triplet heavy lepton $\Sigma_R$ does  not have hypercharge, no $Z$-penguin contribution and therefore $\mu - e$ conversion gives weaker constraint compared with $\mu \to e \gamma$. From the figure, we see that the quadruplet Yukawa couplings are constrained by the new MEG data to be less than $0.1$ for a wide range of parameter space. As we showed $y_\nu$ is typically $10^{-6}$ for $1$ TeV quadruplet scalars in both normal and inverted neutrino mass spectrum. The contribution from $Y_\nu$ is negligibly small. On the other hand, $y_\chi$ can be enhanced by a factor of $1/\sqrt{\kappa} \simeq 12\pi/\sqrt{-\lambda_5}$ with $m_N \sim m_\chi \sim 1$TeV. To obtain $y_\chi \simeq 0.1$, $\lambda_5 \simeq 10^{-8}$ is required. Such a tiny $\lambda_5$ can be naturally understood as a remnant of the lepton number symmetry. The quadruplet model can have Yukawa coupling producing $\mu \to e \gamma$ closing to the present upper bound. Improved experimental limits can further constrain the model parameters.

In the right panel of FIG.\ref{FIG:LFV}, we also show the future prospects of LFV bounds. For $\mu \to e\gamma$ we take $B(\mu \to e\gamma)=1\times 10^{-13}$\cite{Ref:MEG-pot} as the near future improved MEG experimental sensitivity. For $\mu - e$ conversion, there are several planed new experiments, such as Mu2E\cite{Ref:Mu2E}/COMET\cite{Ref:COMET} and PRISM\cite{Ref:PRISM} for $\mu -e $ conversion using Al and Ti. The sensitivities are expected to reach $10^{-16}$\cite{Ref:COMET} and $10^{-18}$\cite{Ref:PRISM}, respectively.
For Ti and Al nuclei, the relevant parameters for our calculations are given by $D(\text{Ti})=0.0864$, $V^{(p)}(\text{Ti})=0.0396, V^{(n)}(\text{Ti})=0.0468$ and $R^0_{\mu\to e}(\text{Ti})=0.0041$, and $D(\text{Al})=0.0362$, $V^{(p)}(\text{Al})=0.0161, V^{(n)}(\text{Al})=0.0173$ and $R^0_{\mu\to e}(\text{Al})=0.0026$\cite{Ref:Mu2Ecv1}.
We see that improved $\mu \to e\gamma$ and $\mu-e$ conversion experiments can further constrain  the quadruplet Yukawa coupling constant. Also note that searches for $\mu- e$ conversions can provide better constraints than that for $\mu \to e \gamma$.

\subsection{Collider signatures of doubly charged Higgs bosons in quadruplet}

Finally, we would like to make some comments about collider aspects of this model. One of the interesting feature of Type III seesaw is that the heavy leptons with a mass of a TeV or lower can be produced at the LHC. The collider phenomenology related to Type III seesaw for the heavy leptons has been studied in great detail\cite{Ref:type3lhc,Ref:type3lhc2}. The introduction of quadruplet also leads to new phenomena in collider physics.

An interesting feature is the existence of the doubly charged particle $\chi^{++}$ in the model. Doubly charged scalar bosons also appear in other models for neutrino masses, for example, Higgs triplet in Type II seesaw Model\cite{Ref:TypeII-1,Ref:TypeII-2}, and Zee-Babu model\cite{Ref:ZeeBabu}.
The doubly charged scalar bosons can be produced at a hadron collider through the Drell-Yan production mechanism $q\bar q\to \gamma, (Z^*) \to \chi^{++}\chi^{--}$\cite{Ref:th-H++1,Ref:th-H++2}.
The vector boson fusion mechanism can also be useful to produce doubly charged particle\cite{Ref:VBF-H++} if the VEV's of the Higgs triplet $v_\Delta^{}$ and the quadruplet $v_\chi$ are not very small.
The recently results from LHC exclude doubly charged Higgs mass to be around $150$ GeV if it decay predominantly through leptonic decay\cite{Ref:LHC-H++}.
Unlike the Type II seesaw and Zee-Babu models, the quadruplet scalars do not have direct interaction with a pair of SM fermion and therefore cannot decay into them.
The lower limit on the mass of doubly charged Higgs boson does not apply for our model.

In both Type II seesaw and the quadruplet models, if the VEV's $v_\Delta^{}$ and $v_\chi$ are not very small, the doubly charged scalar
will mainly decay into a pair of $W^\pm W^\pm$\cite{Ref:type2lhc}. Zee-Babu model does not have such decay modes. In the case of Type II seesaw model, if $v_\Delta < 10^{-4}$ GeV, the leptonic pair decay modes will become the dominate one for the doubly charged scalar because the decay to gauge boson pair is suppressed by $v_\Delta$ while leptonic Yukawa coupling is scaled as $m_\nu/v_\Delta$. This is, however, not the case for quadruplet model.

%

The $\chi^{++}$ can couple to $e^+ \Sigma^+$ through Yukawa coupling. Since the heavy charged lepton $\Sigma^+$ can mixing with $e^+$ because mixing in eq. \eqref{mixing} leading to $\chi^{++} \to e^+ e^+$. However, the mixing in this case is proportional to $Y_\nu^2v^2/2m^2_\Sigma$ which is small. There is another possible decay for $\chi^{++}$. Electromagnetic loop correction\cite{Ref:MinDM} will make $\Sigma^+$ to be heavier than $\Sigma^0$ allowing $\Sigma^+ \to \pi^+ \Sigma^0$.
Then $\Sigma^0$ mixes with light neutrinos to allow $\Sigma^0 \to e^\pm W^\mp$ decay. Since the mixing between light neutrino and $\Sigma^0$ is only suppressed by a factor $Y_\nu v/m_\Sigma$, the decay mode, $\chi^{++} \to e^+ \pi^+ e^\pm W^\mp$ would be more important than $\chi^{++}\to e^+ e^+$. This is different than Type II seesaw model in the case even the VEV's are very small.

\section{Conclusions}
In Type III seesaw the heavy neutrinos are contained in leptonic triplet representations. Being a triplet of $SU(2)_L$ gauge group, the heavy leptons have non-trivial structure. Concerning Yukawa interaction for seesaw mechanism, we find a new possibility of having new type of Yukawa couplings by introducing a quadruplet $\chi$ with hypercharge equal to half. When the neutral component field of $\chi$ develops a non-zero VEV, a Dirac mass terms connecting the light and heavy neutrinos can result to facilitate the seesaw mechanism.  It is interesting to note that the VEV of the quadruplet Higgs is constrained to be very small from electroweak precision data. Therefore the Yukawa couplings of a quadruplet can be much larger than those in a Type III model with a Higgs doublet only.
We also find that unlike the usual Type III seesaw model where at least two copies of leptonic triplets are needed, with both doublet and quadruplet Higgs representations, just one leptonic triplet is possible to have a phenomenologically acceptable model because light neutrino masses can receive sizable contributions from both the tree and one loop levels. Large Yukawa coupling may have observable effects on lepton flavor violating processes, such as  $\mu \to e \gamma$ and $\mu - e$ conversion. There are also some interesting collider signatures for the doubly charged particle in the quadruplet model.
\\

{\bf Acknowledgment}:

\noindent We would like to thank Dr. Sugiyama for useful discussions. This work was partially supported by \#20540282 and \#21104004, NSC and NCTS of ROC, and SJTU 985 grant of China.


\end{document}